# Surprisingly Good Performance of XYG3 Family Functionals Using Scaled KS-MP3 Correlation

*Golokesh Santra, Emmanouil Semidalas, and Jan M.L. Martin*\*

Department of Molecular Chemistry and Materials Science, Weizmann Institute of Science, 7610001 Reḥovot, Israel.

Email: gershom@weizmann.ac.il

**Abstract:** By adding a GLPT3 (third-order Görling-Levy perturbation theory, or KS-MP3) term $E_3$ to the XYG7 form for a double hybrid, we are able to bring down WTMAD2 (weighted total mean absolute deviation) for the very large and chemically diverse GMTKN55 benchmark to an unprecedented 1.17 kcal/mol, competitive with much costlier composite wavefunction ab initio approaches. Intriguingly: (a) the introduction of $E_3$ makes an empirical dispersion correction redundant; (b) GGA or mGGA semilocal correlation functionals offer no advantage over LDA in this framework; (c) if a dispersion correction is retained, then simple Slater exchange leads to no significant loss in accuracy. It is possible to create a 6-parameter functional with WTMAD2=1.42 that has no post-LDA DFT components and no dispersion correction in the final energy.

Over the past few decades, Kohn-Sham density functional theory[1] (KS-DFT) has become the *go-to* electronic structure method, especially for medium and large molecular systems, as it combines reasonable accuracy with relatively mild computational cost-scaling with system size.[2] Being at the top of the '*Jacob's Ladder*' of density functional approximations,[3] double hybrid density functionals (DHs) (see Refs.[4–8] for reviews) are the most accurate DFT methods available to date for modeling a variety of chemical problems, like main group energetics,[9–11] transition metal catalysis,[12–15] electronic excitation spectroscopy,[16–24] external magnetic[25–28] and electric field[29,30] response properties, and so on. Double hybrids can be further categorized according to their use of lower rung (hybrid) functionals in the orbital generation step as gDHs (after Grimme[31]) and xDHs (after the XYG3[32] and related functionals[5]). Xu and co-workers used B3LYP[33–35] orbitals while calculating the final energy for the original XYG3[32] functional and its later variants, lrc-XYG3[36] and XYGJ-OS.[37] Along this line, xDSD,[38–40] ωDSD,[39] and ωDSD3[41] functionals from Martin group and the combinatorially optimized range separated double hybrid, ωB97M(2)[42] by Mardirossian and Head-Gordon have shown remarkable performance for extensive and chemically diverse main-group benchmark like MGCDB84[9] (Main-Group Chemistry Database with about 5000 test cases divided into 84 subsets) and GMTKN55[10] (General Main-group Thermochemistry, Kinetics, and Noncovalent interactions, 55 problem subsets). In a recent study, Zhang and Xu[43] explored the limits of the XYG3-type double hybrids utilizing B3LYP reference orbitals to be consistent with the original XYG3 functional. By gradually relaxing the constraints of the XYG3 they were able to improve the performance of their new xDHs and with all the constraints removed, XYG7@B3LYP appeared as the best in class. Their most intriguing finding was that decoupling Slater exchange from the GGA enhancement factor, i.e., giving B88x and pure Slater exchange independent coefficients, markedly improved performance.

In this letter, we consider two avenues for potential further improvement over XYG7@B3LYP. First, whether there might be more suitable orbitals than B3LYP for this type of functionals. Second, in light of our previous findings for the DSD3 and ωDSD3 functionals,[41] whether it would be advantageous to add a scaled nonlocal GLPT3 (third-order Görling-Levy[44] perturbation theory) term on top of the same-spin and opposite-spin GLPT2 (second-order GLPT) terms already present. To answer both questions, we begin by employing reference orbitals where the percentage HF-like exchange (HFx) is increased systematically. Following Ref. [43], the number m in XYGm refers to the number of terms considered; as we shall add a D3BJ[45–47] empirical dispersion correction term, we shall refer to these xDH's as XYG8@B$n$LYP and XYG9@B$n$LYP, where $n$ stands for the percentage of HFx used in the orbital generation. The final energy expressions of the XYG8@BnLYP and XYG9@B$n$LYP functionals have the form:

$$E_{XYG8@BnLYP} = E_{N1e} + c_{X,HF}E_{X,HF} + c_{X,GGA}E_{X,B88} + c_{X,LDA}E_{X,Slater} + c_{C,GGA}E_{C,LYP} + c_{C,LDA}E_{C,VWN5} + c_{2ab}E_{2ab} + c_{2ss}E_{2ss} + s_6 E_{disp}$$

$$E_{XYG9@BnLYP} = E_{N1e} + c_{X,HF}E_{X,HF} + c_{X,GGA}E_{X,B88} + c_{X,LDA}E_{X,Slater} + c_{C,GGA}E_{C,LYP} + c_{C,LDA}E_{C,VWN5} + c_{2ab}E_{2ab} + c_{2ss}E_{2ss} + c_3 E_3 + s_6 E_{disp}$$

In both expressions, $E_{N1e}$ stands for the sum of nuclear repulsion and one-electron energy terms; $E_{X,HF}$ represents the Hartree–Fock (HF) like exact exchange and $c_{X,HF}$ the corresponding coefficient; $E_{X,B88}$ and $E_{X,Slater}$ are the exchange energy components from the semilocal Becke88 generalized gradient approximation(GGA) and Slater-type local density approximation (LDA) — $c_{X,GGA}$ and $c_{X,LDA}$ are corresponding parameters respectively. The semilocal GGA and LDA correlation energies are represented by $E_{C,LYP}$ and $E_{C,VWN5}$, where $c_{C,GGA}$ and $c_{C,LDA}$ are the respective coefficients for those energy components. $E_{2ab}$ and $E_{2ss}$ are the opposite-spin and same-spin GLPT2 energies, $E_3$ is the GLPT3 energy, and $c_{2ab}$, $c_{2ss}$, and (for XYG9) $c_3$ are their respective linear coefficients. Finally, $E_{disp}$ is a dispersion correction such as D3(BJ)[45–47] or D4[48–50], each coming with adjustable parameters. In the present work, the dispersion energy term is made proportional to a linear coefficient $s_6$.

In the discussion below, XYG8[$f1$]@B$n$LYP will indicate that one parameter was frozen, XYG8[$f2$]@B$n$LYP that two parameters were frozen, and so on.

Unlike B3LYP, B$n$LYP reference orbitals used for our XYG8 and XYG9 functionals do not contain any "unenhanced" Slater exchange. The B$n$LYP orbitals only differ in the exchange part where $n$% HFx and (100-$n$)% DFTx are used while the correlation part is kept unchanged throughout (19% VWN5 + 81% LYP). During the orbital generation and the final energy evaluation step, we make use of the same semilocal GGA-type and LDA-type exchange and correlation functionals.

For the present study, we have used the aforementioned GMTKN55 benchmark suite[10] throughout to validate and parametrize our new functionals. It comprises 55 types of chemical problems, which can be further classified into five major subcategories: thermochemistry of small and medium-sized molecules, barrier heights, large-molecule reactions, intermolecular interactions, and conformer energies (or intramolecular interactions) (for the details of all 55 subsets with proper references see Table S1 in Supporting Information). The WTMAD2 (so-called weighted total mean absolute deviation of type 2) has been used as our primary metric of choice:

$$\text{WTMAD2} = \frac{1}{\sum_{i=1}^{55} N_i} \cdot \sum_{i=1}^{55} N_i \cdot \frac{56.84 \text{ kcal/mol}}{|\Delta E|_i} \cdot \text{MAD}_i$$

where $\overline{|\Delta E|}_i$ is the mean absolute value of all the reference energies from $i = 1$ to 55, $N_i$ is the number of systems in each subset, $MAD_i$ is the mean absolute difference between calculated and reference energies for each of the 55 subsets.

All electronic structure calculations except the third-order Møller-Plesset correlation (MP3) part were performed using the MRCC2020[51] package. The Weigend–Ahlrichs family def2-QZVPP[52] basis set was used for all subsets except for the anion-containing ones WATER27, RG18, IL16, G21EA, BH76, BH76RC and AHB21 – where the diffuse-function augmented def2-QZVPPD[53] was employed instead – and the C60ISO and UPU23 subsets, where we settled for the def2-TZVPP basis set to reduce computational cost. To reduce the latter further, the corresponding standard RI[54] and RI-JK[55] basis sets were employed for the correlation as well as for the Coulomb (J) and HF exchange (K) parts. The LD0110-LD0590 angular integration grid was used throughout, being a pruned Lebedev-type integration grid similar to `Grid=UltraFine` in Gaussian[56] or SG-3 in Q-Chem.[57] The same frozen core settings were used as in Refs.[40,41] MP3 calculations in both HF and KS orbitals were performed using Q-Chem 5.4,[57] def2-TZVPP being used throughout — we previously found that this basis set is adequate for post-MP2 corrections, both in composite WFT and in double-hybrid DFT contexts.[41,58] All calculations were performed on the ChemFarm HPC cluster in the Faculty of Chemistry at the Weizmann Institute of Science.

A fully optimized XYG8@B$n$LYP functional has eight adjustable parameters: $c_{X,HF}$, $c_{X,GGA}$, $c_{X,LDA}$, $c_{C,GGA}$, $c_{C,LDA}$, $c_{2ab}$, $c_{2ss}$(=$c_{2aa}$+$c_{2bb}$), and for the D3(BJ) dispersion correction, one prefactor $s_6$ and one parameter $a_2$ for the damping function (like in refs[59,60] we constrain $a_1$=$s_8$=0). The XYG9@B$n$LYP functional has one additional prefactor ($c_3$) for the $E_3$ component.

We employed Powell's BOBYQA[61] (Bound Optimization BY Quadratic Approximation) derivative-free constrained optimizer, together with scripts and Fortran programs developed in-house, for the optimization of all parameters.

The statistical significance of adding or omitting a parameter was decided with the help of the Bayesian Information Criterion.[62,63] For the dataset at hand (see Appendix A in the Supporting Information for details), a change of ~0.6% in WTMAD2 corresponds to "strong" and ~3.5% to "decisive" importance according to the Kass-Raftery criteria.[63]

We first explore the performance of the XYG8@B$n$LYP functionals with $n$ ranging from 10-70%. The minimum WTMAD2 occurs near $n$ = 20%; in fact, the WTMAD2 surface is relatively flat between 20-30%. While optimizing all eight parameters, the coefficient for the semilocal GGA-type correlation ($c_{C,GGA}$) settles near *zero*, thus fixing it to zero and reoptimizing the remaining seven parameters led us to WTMAD2=1.85 kcal/mol. That being said, the XYG8[*f1*]@B$_{25}$LYP and XYG8[*f1*]@B$_{30}$LYP are two very close contenders — which illustrates the flat behavior of the WTMAD2 surface between the 20-30% region (see Figure1).

Next, what if we leave out the dispersion component entirely, thus eliminating one additional parameter? On average, this would sacrifice only ~0.08 kcal/mol accuracy in terms of total WTMAD2, or about 4% of the total, which just exceed the "decisive" BIC criterion. As expected, among the top five subcategories of GMTKN55, the noncovalent interactions, and to some extent, barrier heights are most affected (see Table S2 in the Supporting Information). Although the D3BJ-uncorrected XYG8[*f2*]@B$_{25}$LYP functional offers the lowest WTMAD2 (1.92 kcal/mol), the performances of the XYG8[*f2*] variants utilizing B$_{20}$LYP and B$_{30}$LYP orbitals are pretty close.

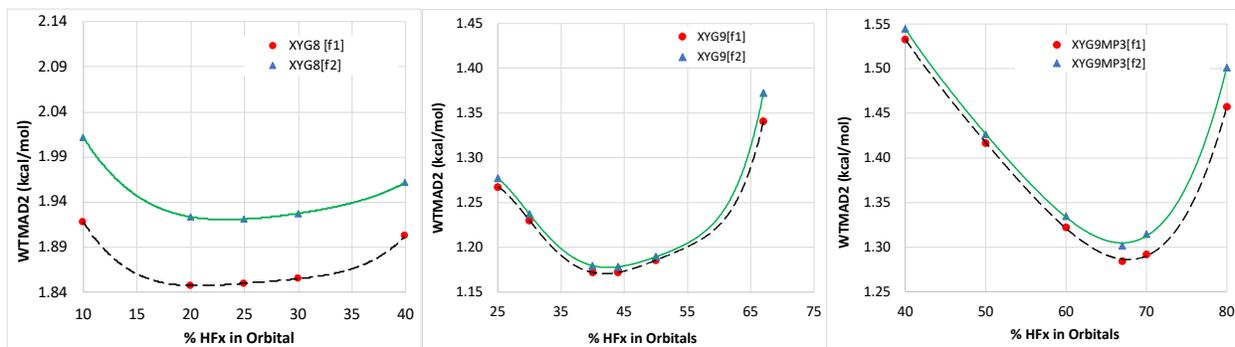

**Figure 1: Effect of different orbitals on the total WTMAD2 (kcal/mol) for the PT2-based (left), KS-MP3 based (center) and canonical MP3 based (right) xDHs.**

For the XYG8[*f1*]@BnLYP functionals, the coefficients for the GGA exchange (i.e., $c_{X,GGA}$) decrease as *n* increases (see Table 1). So, together with $c_{C,GGA}$ if we also fix $c_{X,GGA}$ at zero and optimize the remaining coefficients, the WTMAD2 difference between the XYG8[*f1*]@BnLYP and the new form, XYG8[*f2g*]@BnLYP, becomes narrower with increasing HF-like exchange in the orbitals and almost invisible for n = 70% (see Table 1). The letter *g* in XYG8[*f2g*]@BnLYP refers to the nature of the frozen parameters — *g* for GGA and *l* for LDA. Going from XYG8[*f1*]@B$_{20}$LYP to XYG8[*f2g*]@B$_{20}$LYP WTMAD2 increases by 0.16 kcal/mol. Now, comparing these two for the five top-level subsets, performance for all four top-level subsets other than barrier heights is degraded by imposing the $c_{X,GGA}=c_{C,GGA}=0$ constraint (see Table S2). Interestingly, among the small molecule thermochemistry subsets, the G21EA subset actually improves by imposing that constraint. Now, if we omit the dispersion correction term only and reoptimize everything else (i.e, XYG8[*f1d*]@B$_{20}$LYP, where *d* stands for dispersion), we obtain WTMAD2= 1.92 kcal/mol, which is practically the same as what we got for the XYG8[*f2*]@B$_{20}$LYP functional. Interestingly for XYG8[*f1*]@B$_{20}$LYP, if both $s_6$ and $c_{X,GGA}$ are frozen (i.e, XYG8[*f3*]) this drives up WTMAD2 by almost 1 kcal/mol, an order of magnitude more than freezing each of these individually (see Table S1 in the Supporting Information).

Next, as a control experiment: by fixing the coefficients for the LDA exchange and correlation ($c_{C,LDA}$ and $c_{X,LDA}$) at *zero*, we always obtain higher WTMAD2 than the corresponding functional forms where both $c_{X,GGA}$ and $c_{C,GGA}$ are zero. Once again, the WTMAD2 difference seems to become narrower for larger values of *n* (see Table S2 in the Supporting Information). The best pick in this category, the XYG8[*f2l*]@B$_{50}$LYP offers WTMAD2= 2.12 kcal/mol, which is comparable to our group's latest xDHs, xDSD$_{75}$-PBEP86-D3BJ (WTMAD2 = 2.14 kcal/mol) and xDSD$_{75}$-PBEP86-D4 (WTMAD2 = 2.12 kcal/mol).[39] Incidentally, for XYG8[*f5*]@B20LYP, using only 100% HF-like exchange with scaled LDAc and PT2 correlation, WTMAD2 rises to 3.59 kcal/mol (see Table 1).

We now explore the effect of using different LDA semilocal correlations for the construction of these new functionals. XYG7@B3LYP is chosen for this purpose, where the VWN3 correlation is replaced by VWN5. For B3LYP orbitals, we also checked PW92 and found the results to be essentially the same as for VWN5 (see Table S3 and S4 in the Supporting Information). Without further optimization, the original XYG7[43] offers WTMAD2=1.88 kcal/mol, which is 0.17 kcal/mol lower than the value reported by Zhang and Xu. This discrepancy could be attributed to differences in the basis sets used, particularly for anionic subsets.

Table 1: Total WTMAD2 (kcal/mol) and final parameters for the PT2-based XYG8@B$n$LYP double hybrids.[a]

| Functionals | Parameters | | | | | | | | WTMAD2 (kcal/mol) |
|---|---|---|---|---|---|---|---|---|---|
| | $c_{X,HF}$ | $c_{X,GGA}$ | $c_{X,LDA}$ | $c_{C,GGA}$ | $c_{C,LDA}$ | $c_{2ab}$ | $c_{2ss}$ | $s_6$[b] | |
| XYG8[*f1*]@B$_{10}$LYP | 0.8226 | -0.0929 | 0.2530 | [0] | 0.5627 | 0.3407 | 0.1989 | 0.2457 | 1.918 |
| XYG8[*f2*]@B$_{10}$LYP | 0.8921 | -0.1687 | 0.2447 | [0] | 0.5937 | 0.3488 | 0.2824 | [0] | 2.012 |
| XYG8[*f2g*]@B$_{10}$LYP | 0.7570 | [0] | 0.2421 | [0] | 0.5480 | 0.3133 | 0.1743 | 0.4399 | 2.052 |
| XYG8[*f2l*]@B$_{10}$LYP | 0.8382 | 0.1539 | [0] | 0.6991 | [0] | 0.3185 | 0.2701 | 0.2386 | 2.254 |
| **XYG8[*f1*]@B$_{20}$LYP** | **0.8482** | **-0.0935** | **0.2270** | **[0]** | **0.5243** | **0.3985** | **0.2162** | **0.2060** | **1.848** |
| **XYG8[*f2*]@B$_{20}$LYP** | **0.9085** | **-0.1603** | **0.2236** | **[0]** | **0.5264** | **0.4189** | **0.2780** | **[0]** | **1.924** |
| XYG8[*f2g*]@B$_{20}$LYP | 0.7807 | [0] | 0.2117 | [0] | 0.5436 | 0.3567 | 0.1996 | 0.4206 | 2.003 |
| XYG8[*f2l*]@B$_{20}$LYP | 0.8576 | 0.1373 | [0] | 0.6249 | [0] | 0.3864 | 0.2773 | 0.2072 | 2.179 |
| XYG8[*f5*]@B$_{20}$LYP | [1.0] | [0] | [0] | [0] | 0.6113 | 0.3835 | 0.6174 | [0] | 3.594 |
| XYG8[*f1*]@B$_{25}$LYP | 0.8724 | -0.1071 | 0.2097 | [0] | 0.5299 | 0.4233 | 0.2468 | 0.1550 | 1.850 |
| XYG8[*f2*]@B$_{25}$LYP | 0.9144 | -0.1537 | 0.2132 | [0] | 0.4928 | 0.4540 | 0.2780 | [0] | 1.921 |
| XYG8[*f2g*]@B$_{25}$LYP | 0.7921 | [0] | 0.2033 | [0] | 0.5026 | 0.3978 | 0.1987 | 0.3952 | 1.991 |
| XYG8[*f2l*]@B$_{25}$LYP | 0.8638 | 0.1334 | [0] | 0.5803 | [0] | 0.4202 | 0.2884 | 0.2006 | 2.161 |
| XYG8[*f1*]@B$_{30}$LYP | 0.8674 | -0.0922 | 0.2015 | [0] | 0.5096 | 0.4501 | 0.2397 | 0.1824 | 1.855 |
| XYG8[*f2*]@B$_{30}$LYP | 0.9180 | -0.1458 | 0.1986 | [0] | 0.4964 | 0.4728 | 0.2949 | [0] | 1.928 |
| XYG8[*f2g*]@B$_{30}$LYP | 0.7972 | [0] | 0.1953 | [0] | 0.4992 | 0.4188 | 0.2076 | 0.3846 | 1.987 |
| XYG8[*f2l*]@B$_{30}$LYP | 0.8577 | 0.1409 | [0] | 0.5530 | [0] | 0.4485 | 0.2777 | 0.2351 | 2.146 |
| XYG8[*f1*]@B$_{40}$LYP | 0.8951 | -0.1015 | 0.1754 | [0] | 0.5075 | 0.5004 | 0.2896 | 0.1169 | 1.903 |
| XYG8[*f2*]@B$_{40}$LYP | 0.9229 | -0.1334 | 0.1762 | [0] | 0.4982 | 0.5132 | 0.3244 | [0] | 1.962 |
| XYG8[*f2g*]@B$_{40}$LYP | 0.8090 | [0] | 0.1855 | [0] | 0.4378 | 0.4902 | 0.2034 | 0.3589 | 1.987 |
| XYG8[*f2l*]@B$_{40}$LYP | 0.8555 | 0.1421 | [0] | 0.5139 | [0] | 0.5012 | 0.2797 | 0.2506 | 2.129 |
| XYG8[*f1*]@B$_{50}$LYP | 0.8712 | -0.0702 | 0.1738 | [0] | 0.4555 | 0.5519 | 0.2595 | 0.1856 | 1.935 |
| XYG8[*f2*]@B$_{50}$LYP | 0.9269 | -0.1227 | 0.1608 | [0] | 0.4661 | 0.5779 | 0.3326 | [0] | 2.005 |
| XYG8[*f2g*]@B$_{50}$LYP | 0.8199 | [0] | 0.1681 | [0] | 0.4344 | 0.5296 | 0.2457 | 0.3315 | 1.988 |
| XYG8[*f2l*]@B$_{50}$LYP | 0.8543 | 0.1401 | [0] | 0.4951 | [0] | 0.5555 | 0.2756 | 0.2535 | 2.119 |
| XYG8[*f1*]@B$_{60}$LYP | 0.8753 | -0.0684 | 0.1632 | [0] | 0.4411 | 0.5933 | 0.2931 | 0.1696 | 1.997 |
| XYG8[*f2*]@B$_{60}$LYP | 0.9197 | -0.1156 | 0.1593 | [0] | 0.4281 | 0.6269 | 0.3417 | [0] | 2.080 |
| XYG8[*f2g*]@B$_{60}$LYP | 0.8146 | [0] | 0.1726 | [0] | 0.4022 | 0.5861 | 0.2245 | 0.3328 | 2.006 |
| XYG8[*f2l*]@B$_{60}$LYP | 0.8623 | 0.1277 | [0] | 0.4655 | [0] | 0.6096 | 0.3062 | 0.2110 | 2.129 |
| XYG8[*f1*]@B$_{70}$LYP | 0.8413 | -0.0318 | 0.1631 | [0] | 0.4294 | 0.6213 | 0.2803 | 0.2548 | 2.106 |
| XYG8[*f2*]@B$_{70}$LYP | 0.9167 | -0.1014 | 0.1442 | [0] | 0.4488 | 0.6572 | 0.3803 | [0] | 2.227 |
| XYG8[*f2g*]@B$_{70}$LYP | 0.8142 | [0] | 0.1682 | [0] | 0.4026 | 0.6250 | 0.2431 | 0.3227 | 2.108 |
| XYG8[*f2l*]@B$_{70}$LYP | 0.8222 | 0.1675 | [0] | 0.4796 | [0] | 0.6258 | 0.2821 | 0.3118 | 2.200 |

[a]parameters in square bracket are the constraints we used. [b]$a_2$ is kept constant at 5.6 throughout.

In 2003, Grimme introduced the SCS-MP3 method, which outperformed regular MP3 at no extra cost;[64] later, Hobza and coworkers[65] coined the term MP2.5 for the average of MP2 and MP3, and showed its superior performance for noncovalent interactions. Recently, we have reported[41] that adding an MP3-like correlation term in the final KS energy expression can significantly improve the performances of both global and range-separated DSD double hybrids, albeit at some extra computational expense. As technical limitations precluded using KS reference orbitals for MP3 at that time, we used the HF reference orbitals to calculate the conventional MP3 correlation energies and added them into the DSD and ωDSD functional forms with an additional linear coefficient. Motivated by the above two findings, here we combine the scaled $E_3$ correlation energies with the XYG8@BnLYP functionals to check whether it is possible to improve their performance further. Similar to Ref.[41], we have omitted the 50 largest systems, i.e., the UPU23 and C60 subsets, ten largest ISOL24, three INV24, and a single IDISP, we refer to this reduced version of GMTKN55 as mod-GMTKN55 in this paper.

Surprisingly enough, the lowest WTMAD2 we got is 1.28 kcal/mol using the $B_{67}$LYP orbitals. Similar to the PT2 based functionals above, here too, the optimized $c_{C,GGA}$ values approach *zero*, and thus $c_{C,GGA}$ can be eliminated without any accuracy loss (see Table 2). Upon comparing the performance of the eight parameter XYG9$_{MP3}$[*f1*]@$B_{67}$LYP with the PT2 based XYG8[*f1*]@$B_{20}$LYP for the five top-level subsets, we find that the lion's share of the WTMAD2 improvement is seen in the small and medium molecule thermochemistry and large molecule reactions subsets. Next, comparing the XYG9$_{MP3}$[*f1*] with the PT2-based XYG8[*f1*] evaluated using the same set of orbitals, namely $B_{67}$LYP, we find that statistics of large-molecule reactions subset improve by 50% upon inclusion of MP3 correlation (see Table S5 in the Supporting Information).

**Table 2: Total WTMAD2 (kcal/mol) and final parameters for the XYG9$_{MP3}$@B*n*LYP double hybrids.[a]**

| Functionals | Parameters | | | | | | | | | WTMAD2 (kcal/mol) |
|---|---|---|---|---|---|---|---|---|---|---|
| | $c_{X,HF}$ | $c_{X,GGA}$ | $c_{X,LDA}$ | $c_{C,GGA}$ | $c_{C,LDA}$ | $c_{2ab}$ | $c_{2aa}$ | $c_3$ | $s_6$[b] | |
| XYG9$_{MP3}$[*f1*]@$B_{10}$LYP | 0.8784 | -0.1297 | 0.2341 | [0] | 0.5176 | 0.3704 | 0.2917 | 0.1574 | 0.1349 | 1.783 |
| XYG9$_{MP3}$[*f2*]@$B_{10}$LYP | 0.9180 | -0.1675 | 0.2237 | [0] | 0.5433 | 0.3742 | 0.3465 | 0.1702 | [0] | 1.838 |
| XYG9$_{MP3}$[*f1*]@$B_{20}$LYP | 0.8862 | -0.1073 | 0.2060 | [0] | 0.4718 | 0.4306 | 0.3002 | 0.1591 | 0.1548 | 1.671 |
| XYG9$_{MP3}$[*f2*]@$B_{20}$LYP | 0.9392 | -0.1499 | 0.1880 | [0] | 0.4733 | 0.4449 | 0.3872 | 0.2120 | [0] | 1.694 |
| XYG9$_{MP3}$[*f1*]@$B_{25}$LYP | 0.9309 | -0.1197 | 0.1722 | [0] | 0.4356 | 0.4758 | 0.3894 | 0.2290 | 0.0612 | 1.641 |
| XYG9$_{MP3}$[*f2*]@$B_{25}$LYP | 0.9454 | -0.1405 | 0.1757 | [0] | 0.4320 | 0.4851 | 0.3936 | 0.2178 | [0] | 1.659 |
| XYG9$_{MP3}$[*f1*]@$B_{30}$LYP | 0.9395 | -0.1080 | 0.1504 | [0] | 0.4220 | 0.5033 | 0.4322 | 0.2624 | 0.0617 | 1.609 |
| XYG9$_{MP3}$[*f2*]@$B_{30}$LYP | 0.9538 | -0.1318 | 0.1542 | [0] | 0.4367 | 0.5037 | 0.4437 | 0.2595 | [0] | 1.625 |
| XYG9$_{MP3}$[*f1*]@$B_{40}$LYP | 0.9387 | -0.0875 | 0.1324 | [0] | 0.3744 | 0.5726 | 0.4398 | 0.2667 | 0.0851 | 1.533 |
| XYG9$_{MP3}$[*f2*]@$B_{40}$LYP | 0.9617 | -0.1087 | 0.1299 | [0] | 0.3530 | 0.5914 | 0.4695 | 0.2845 | [0] | 1.545 |
| XYG9$_{MP3}$[*f1*]@$B_{50}$LYP | 0.9627 | -0.0659 | 0.0901 | [0] | 0.2888 | 0.6657 | 0.5718 | 0.4224 | 0.0568 | 1.417 |
| XYG9$_{MP3}$[*f2*]@$B_{50}$LYP | 0.9786 | -0.0831 | 0.0918 | [0] | 0.2684 | 0.6867 | 0.5638 | 0.3986 | [0] | 1.426 |
| XYG9$_{MP3}$[*f1*]@$B_{60}$LYP | 0.9582 | -0.0500 | 0.0794 | [0] | 0.2395 | 0.7424 | 0.6019 | 0.4554 | 0.0649 | 1.322 |
| XYG9$_{MP3}$[*f2*]@$B_{60}$LYP | 0.9789 | -0.0625 | 0.0680 | [0] | 0.2492 | 0.7512 | 0.6419 | 0.4653 | [0] | 1.335 |
| **XYG9$_{MP3}$[*f1*]@$B_{67}$LYP** | **0.9490** | **-0.0182** | **0.0606** | **[0]** | **0.1907** | **0.8002** | **0.6676** | **0.5381** | **0.0962** | **1.284** |
| **XYG9$_{MP3}$[*f2*]@$B_{67}$LYP** | **0.9807** | **-0.0408** | **0.0478** | **[0]** | **0.1935** | **0.8165** | **0.7218** | **0.5557** | **[0]** | **1.301** |
| XYG9$_{MP3}$[*f2g*]@$B_{67}$LYP | 0.9391 | [0] | 0.0548 | [0] | 0.1927 | 0.7902 | 0.6818 | 0.5491 | 0.1269 | 1.297 |
| XYG9$_{MP3}$[*f3*]@$B_{67}$LYP | 0.9834 | [0] | 0.0121 | [0] | 0.1572 | 0.8291 | 0.8320 | 0.6478 | [0] | 1.378 |
| XYG9$_{MP3}$[*f1*]@$B_{70}$LYP | 0.9360 | -0.0029 | 0.0589 | [0] | 0.1880 | 0.8089 | 0.6796 | 0.5515 | 0.1269 | 1.292 |
| XYG9$_{MP3}$[*f2*]@$B_{70}$LYP | 0.9778 | -0.0367 | 0.0426 | [0] | 0.2161 | 0.8207 | 0.7527 | 0.5678 | [0] | 1.315 |
| XYG9$_{MP3}$[*f1*]@$B_{80}$LYP | 0.9088 | 0.0241 | 0.0577 | [0] | 0.1830 | 0.8517 | 0.7173 | 0.5751 | 0.1606 | 1.457 |
| XYG9$_{MP3}$[*f2*]@$B_{80}$LYP | 0.9673 | -0.0157 | 0.0365 | [0] | 0.1376 | 0.9242 | 0.8108 | 0.6593 | [0] | 1.501 |

[a]parameters in square brackets are the constraints we used. [b]$a_2$ is kept constant at 5.6 throughout.

In Ref.[41], for none for the global or range-separated DSD3 functionals could we omit the dispersion correction term with impunity: for instance, ωDSD3-PBEP86-D3BJ offered WTMAD2 = 2.08 kcal/mol while the dispersion-free variant ωDSD3-PBEP86 has WTMAD2 = 2.86 kcal/mol (see Table 3 in Ref.[41]). However, for the present conventional-MP3-based XYG9$_{MP3}$[*f1*]@B*n*LYP functionals, discarding the dispersion term has only a negligible effect on total WTMAD2 (~0.01-0.02 kcal/mol) — which is a much narrower range compared to that in PT2-based XYG8[*f1*]@B*n*LYP functionals (see Figures 1).

We now examine the importance of considering the semilocal DFT exchange coefficient ($c_{X,GGA}$). The optimized $c_{X,GGA}$ values decrease as *n* increases (see Table 2). If we fix both $c_{X,GGA}$ and $c_{C,GGA}$ at zero and optimize other coefficients (i.e., XYG9$_{MP3}$[*f2g*]@$B_{67}$LYP), the resulted WTMAD2 goes up only a negligible amount compared to that for the XYG9$_{MP3}$[*f1*]@$B_{67}$LYP. Now, if we also discard the empirical dispersion correction term — which leaves us with only six adjustable parameters (same as our group's revDSD-PBEP86-D3BJ[40]), we get WTMAD2 = 1.38 kcal/mol. So, for these PT3-containing double hybrids, the dispersion correction term is nearly redundant as far as total WTMAD2 is concerned.

Thus far, we have used canonical MP3 correlation energies as an additional component to improve the performance of GLPT2 based XYG8[*f1*]@B*n*LYP double hybrids. In a recent study, Head-Gordon and co-workers have shown[66,67] that the use of DFT orbitals instead of pure HF ones for MP3 calculation can significantly improve performance for thermochemistry, barrier heights, noncovalent interactions, and dipole moments. Additionaly, Jana et al.[68] have found that considering Görling–Levy perturbation terms beyond second order can improve the performance of PBE-based double hybrids for some selected datasets. So, here we explore whether employing the scaled MP3 correlation energies evaluated *in a basis of BnLYP orbitals (i.e., KS-MP3)* can still be beneficial compared to the already excellent XYG9$_{MP3}$@B*n*LYP functionals. We have used the same mod-GMTKN55 dataset for the parametrization as well as validation of these functionals.

We again notice that $c_{C,GGA}$ settles near zero after optimization, so we proceed by setting it to *zero* and refitting the eight remaining parameters for the KS-MP3 based functionals (namely XYG9[*f1*]@B*n*LYP). Using 44% HFx in the orbital generation step, the XYG9[*f1*]@B$_{44}$LYP offers the lowest WTMAD2 (1.17 kcal/mol), which is even lower than what we got for the top canonical MP3 based xDH, XYG9$_{MP3}$[*f1*]@B$_{67}$LYP (1.28 kcal/mol). However, the WTMAD2 surface for XYG9[*f1*]@B*n*LYP is pretty flat between the 40-50% region (see Figure 1). That being said, the performance of the XYG9[*f1*]@B$_{44}$LYP is now in the territory of the best G4-type composite WFT methods (cWFT), recently proposed by Semidalas and Martin.[58,69] It actually surpasses the performance of the G4-T approach[58] by 0.34 kcal/mol; the improvement lies mainly in conformers and to a lesser degree in large molecule energetics (see Table 1 in Ref.[58]). The correlation consistent cc-G4-type protocols[69] still have an edge over XYG9[*f1*]@B$_{44}$LYP, but the accuracy gap has narrowed even more since the top cWFT performers, cc-G4-T and the parameter-free cc-G4-F12-T, have WTMAD2 = 0.90 and 1.04 kcal/mol, respectively.

Now, comparing the XYG9[*f1*]@B$_{44}$LYP with the PT2-based XYG8[*f1*]@B$_{20}$LYP (WTMAD2 = 1.79 kcal/mol for mod-GMTKN55) for the five top-level subsets, we found that the large molecule reactions and small molecule thermochemistry are the two most benefited classes. Let's check where exactly KS-MP3 based functionals benefit over the canonical MP3 based variants. We compare the WTMAD2 contribution of individual subsets between the XYG9[*f1*] and XYG9$_{MP3}$[*f1*] (WTMAD2=1.49 kcal/mol) functionals evaluated over the same set of KS orbitals, B$_{44}$LYP (see Table S12 and S17 in the Supporting Information). Although performance improved for most of the subsets, the top affected subsets are: TAUT15, RSE43, HAL59, PAREL, MCONF, G21EA and CDIE20.

Now, what if we leave the dispersion correction out entirely? The WTMAD2 gap between the XYG9[*f1*]@B*n*LYP and XYG9[*f2*]@B*n*LYP then almost vanishes (see Figure 1). Due to the negligible effect of the dispersion correction term, we elected not to fit the more advanced D4[49,50] dispersion correction on top of XYG9[*f2*]@B*n*LYP. At a reviewer's request, a plot of the interaction energy of stretched Ar$_2$ can be found in the Supporting Information (see Figure S2): proper asymptotic $R^{-6}$ dependence is observed there, indicating this was not sacrificed by omitting the dispersion correction.

Similar to the PT2 based XYG8@B$n$LYP and canonical MP3 based XYG9$_{MP3}$@B$n$LYP double hybrids, $c_{X,GGA}$ gradually decreases with the increase of $n$ (see Table 3). Now, if we fix both $c_{X,GGA}$ and $c_{C,GGA}$ at zero and reoptimize the remain seven parameters, we obtain WTMAD2 = 1.25 kcal/mol. Additionally, if we also discard the empirical dispersion term (WTMAD2 = 1.42 kcal/mol), we lose a nontrivial amount (0.25 kcal/mol) of accuracy compared to our best pick XYG9[*f1*]@B$_{44}$LYP. In contrast, the loss in WTMAD2 is 0.17 kcal/mol compared to the corresponding D3BJ-uncorrected variant XYG9[*f2*]@B$_{44}$LYP.

Next, if we consider only pure HF-exchange (i.e., fixing $c_{X,HF}$=1) and optimize the four remaining correlation prefactors, we get WTMAD2 = 1.56 kcal/mol for the XYG9[*f5*]@B$_{44}$LYP (see Table 3). Comparing XYG9[*f5*]@B$_{44}$LYP with XYG9[*f2*]@B$n$LYP for all 55 subsets of GMTKN55, we find the greatest differences for BUT14DIOL, G21EA, RG18, and BSR36. However, we must mention that the resulting functional XYG9[*f5*]@B$_{44}$LYP is no longer a "double hybrid" in the usual sense, as we are mixing the semilocal and nonlocal terms only for the correlation part. However, if we additionally constrain $c_{C,LDA}$=0 for XYG9[*f5*]@B$_{44}$LYP (ak.a, XYG9[*f6*]@B$_{44}$LYP), WTMAD2 increases to 2.50 kcal/mol.

**Table 3: Total WTMAD2 (kcal/mol) and final parameters for the KS-MP3 based XYG9@B$n$LYP double hybrids.[a]**

| Functionals | Parameters | | | | | | | | | WTMAD2 (kcal/mol) |
|---|---|---|---|---|---|---|---|---|---|---|
| | $c_{X,HF}$ | $c_{X,GGA}$ | $c_{X,LDA}$ | $c_{C,GGA}$ | $c_{C,LDA}$ | $c_{2ab}$ | $c_{2ss}$ | $c_3$ | $s_6$[b] | |
| XYG9[*f1*]@B$_{10}$LYP | 0.9004 | -0.1126 | 0.2028 | [0] | 0.4798 | 0.3751 | 0.4227 | 0.0918 | 0.0571 | 1.578 |
| XYG9[*f2*]@B$_{10}$LYP | 0.9154 | -0.1319 | 0.2045 | [0] | 0.4763 | 0.3831 | 0.4361 | 0.0941 | [0] | 1.602 |
| XYG9[*f1*]@B$_{20}$LYP | 0.9276 | -0.1025 | 0.1662 | [0] | 0.4179 | 0.4532 | 0.4760 | 0.1339 | 0.0408 | 1.302 |
| XYG9[*f2*]@B$_{20}$LYP | 0.9396 | -0.1116 | 0.1619 | [0] | 0.4180 | 0.4558 | 0.5010 | 0.1389 | [0] | 1.316 |
| XYG9[*f1*]@B$_{25}$LYP | 0.9348 | -0.0884 | 0.1450 | [0] | 0.3930 | 0.4868 | 0.5132 | 0.1588 | 0.0479 | 1.267 |
| XYG9[*f2*]@B$_{25}$LYP | 0.9440 | -0.1020 | 0.1472 | [0] | 0.3982 | 0.4902 | 0.5192 | 0.1570 | [0] | 1.277 |
| XYG9[*f1*]@B$_{30}$LYP | 0.9358 | -0.0793 | 0.1343 | [0] | 0.3749 | 0.5196 | 0.5186 | 0.1699 | 0.0580 | 1.230 |
| XYG9[*f2*]@B$_{30}$LYP | 0.9520 | -0.0917 | 0.1302 | [0] | 0.3626 | 0.5305 | 0.5463 | 0.1791 | [0] | 1.237 |
| XYG9[*f1*]@B$_{40}$LYP | 0.9571 | -0.0738 | 0.1098 | [0] | 0.2948 | 0.6157 | 0.5684 | 0.2249 | 0.0241 | 1.172 |
| XYG9[*f2*]@B$_{40}$LYP | 0.9602 | -0.0804 | 0.1123 | [0] | 0.2974 | 0.6156 | 0.5640 | 0.2165 | [0] | 1.179 |
| **XYG9[*f1*]@B$_{44}$LYP** | **0.9508** | **-0.0502** | **0.0928** | **[0]** | **0.2880** | **0.6280** | **0.6144** | **0.2570** | **0.0579** | **1.172** |
| **XYG9[*f2*]@B$_{44}$LYP** | **0.9692** | **-0.0644** | **0.0881** | **[0]** | **0.2730** | **0.6447** | **0.6467** | **0.2761** | **[0]** | **1.178** |
| XYG9[*2g*]@B$_{44}$LYP | 0.9165 | [0] | 0.0844 | [0] | 0.2836 | 0.6112 | 0.6048 | 0.2606 | 0.1687 | 1.252 |
| XYG9[*f3*]@B$_{44}$LYP | 0.9652 | [0] | 0.0430 | [0] | 0.2267 | 0.6564 | 0.7641 | 0.3303 | [0] | 1.424 |
| XYG9[*f5*]@B$_{44}$LYP | [1.0] | [0] | [0] | [0] | 0.2703 | 0.6535 | 0.8475 | 0.3488 | [0] | 1.557 |
| XYG9[*f6*]@B$_{44}$LYP | [1.0] | [0] | [0] | [0] | [0] | 0.6986 | 0.9857 | 0.4608 | [0] | 2.499 |
| XYG9[*f1*]@B$_{50}$LYP | 0.9654 | -0.0529 | 0.0782 | [0] | 0.2631 | 0.6771 | 0.6675 | 0.3026 | 0.0173 | 1.186 |
| XYG9[*f2*]@B$_{50}$LYP | 0.9693 | -0.0571 | 0.0785 | [0] | 0.2592 | 0.6809 | 0.6723 | 0.3047 | [0] | 1.189 |
| XYG9[*f1*]@B$_{67}$LYP | 0.9389 | -0.0172 | 0.0674 | [0] | 0.2161 | 0.7795 | 0.6974 | 0.3912 | 0.0851 | 1.341 |
| XYG9[*f2*]@B$_{67}$LYP | 0.9667 | -0.0403 | 0.0610 | [0] | 0.2113 | 0.8080 | 0.7144 | 0.3894 | [0] | 1.372 |

[a]parameters in square brackets are the constraints we used. [b]$a_2$ is kept constant at 5.6 throughout.

For the XYG9[*f1*]@B$_{44}$LYP functionals, we note that the sum of three exchange parameters is very close to one. So, if we re-enforce the $c_{X,HF}$ + $c_{X,GGA}$ + $c_{X,LDA}$ =1 constraint (one among the four constraints of the original XYG3[32]) and optimize the seven remaining parameters, how much performance is sacrificed? Performing this test on both the XYG9[*f1*]@B$_{40}$LYP and XYG9[*f1*]@B$_{44}$LYP and their D3BJ-uncorrected variants, we found that the loss in terms of WTMAD2 is almost imperceptible (see Table S3 in the Supporting Information). So, it is possible to reduce the number of parameters for both variants without sacrificing any significant accuracy.

A reviewer inquired about the counterintuitively negative $c_{X,GGA}$ coefficients in Table 3. In response, we carried out constrained reoptimizations for a sequence of fixed values of $c_{X,HF}$. The result can be found in the Supporting Information (see Figure S3). It can be seen there that $c_{X,GGA}$ has a linear dependence with a negative slope on $c_{X,HF}$; beyond ca. 87%, $c_{X,GGA}$ becomes negative. A tentative explanation can be advanced as follows: considering that $E_{X,B88}=E_{X,LDA} \cdot F_{B88}(s)$, where $F(s)$ is an enhancement factor depending on the reduced density gradient $s=\nabla\rho/\rho^{4/3}$, it follows that $c_{X,LDA} \cdot E_{X,LDA} + c_{X,GGA} \cdot E_{X,B88} = (c_{X,LDA}+c_{X,GGA})E_{X,LDA} + c_{X,GGA}(F_{B88}(s) - 1)$. Now as $F_{B88}(0)=1$ and $F(s) \geq 1$ for all $s$, a negative $c_{X,GGA}$ effectively implies that semilocal exchange will be progressively throttled as $s$ becomes large (in the periphery of the atomic density), while a positive $c_{X,GGA}$ implies the opposite. It stands to reason that, as more "exact" exchange is introduced, the functional would increasingly benefit from semilocal exchange "getting out of the way" as $s$ increases.

Lastly, by adding a KS-MP3 correlation component on top of XYG3@B3LYP and optimizing all four parameters against mod-GMTKN55, we get WTMAD2= 1.64 kcal/mol. On the other hand, doing the same for the XYG7@B3LYP offers WTMAD2= 1.35 kcal/mol (see Table S6 in the Supporting Information).

We can conclude the following from an extensive study of XYG8@B*n*LYP, XYG9$_{MP3}$@B*n*LYP and XYG9@BnLYP double-hybrid density functionals using the large and chemically diverse GMTKN55 and mod-GMTKN55 databases:

- For all three xDH variants we have discussed above, the optimized parameter for the semilocal DFT correlation is close to zero across the board and can be safely eliminated without sacrificing any accuracy in terms of WTMAD2.
- For the new PT2-based XYG8[*f1*]@B*n*LYP double hybrids, we found a minimum WTMAD2 when the final energies are evaluated over B$_{20}$LYP orbitals. However, the WTMAD2 surface is nearly flat over the 20-30% region.
- Including a scaled canonical MP3 term in the final energy significantly improves the performance, as previously found[41] with the DSD3 and ωDSD3 functionals. The XYG9$_{MP3}$[*f1*]@B$_{67}$LYP now offers the lowest WTMAD2 (1.28 kcal/mol) among all the regular MP3 based XYG-type functionals tested above. Among the five top-level subsets, small-molecule thermochemistry and large molecule reactions are the two most benefited classes.
- Replacing the HF-based MP3 term by KS-MP3 further improved statistics, with XYG9[*f1*]@B$_{44}$LYP offering the lowest WTMAD2 (1.17 kcal/mol). To the authors' knowledge, this is by far *the lowest reported WTMAD2 for any DFT method, and is in the same range as the G4-type composite WFT methods.* The WTMAD2 surface is almost flat around the minimum, in the n=40-50% region.
- Another advantage of adding a scaled $E_3$ correlation term in xDHs is that the dispersion correction becomes essentially redundant, unlike the case where $E_3$ is absent. As far as the D3BJ-uncorrected functionals are concerned, KS-MP3 based XYG9[*f2*]@B$_{44}$LYP is our best pick.
- Both for the PT2 based and MP3 based xDHs presented in this letter, the coefficient for the semilocal DFT exchange term decreases with increasing % HFx used during the orbital generation. Thus, this parameter can also be set to zero if one use orbitals with higher

- fraction of HFx for the final energy evaluation. We note that this leads to a situation where the final energy expression is entirely independent of post-LDA enhancement factors in both exchange and correlation.
- Interestingly enough, using the full HF-like exchange and only the four correlation energy terms (i.e., $c_{C,LDA}$, $c_{2ab}$, $c_{2ss}$ and $c_3$) in the final energy expression, we can achieve WTMAD2 = 1.56 kcal/mol for XYG9[*f5*]@B$_{44}$LYP, which is only 0.38 kcal/mol inferior to our best pick XYG9[*f1*]@B$_{44}$LYP.

A final remark is due on the computational cost. Table S18 and Figure S1 show wall clock time dependence on the system size in the ADIM6 subset (n-alkane dimers, n=2–7). XYG8[*f1*]@B$_{44}$LYP and other MP2-based double hybrids, in this size range and with the RI approximation, scale about $O(N^3)$ with basis set size. The RI-MP3 algorithm used scales almost like canonical $O(N^6)$, and hence that term rapidly dominates the overall calculation. However, recent developments[67,70] in tensor hypercontraction approaches[71] reduce scaling to $O(N^4)$ and may be effectively exploited here.

We would like to acknowledge helpful discussions with Prof. Dr. A. Daniel Boese (Karl-Franzens-Universität Graz, Austria). GS acknowledges a doctoral fellowship from the Feinberg Graduate School (WIS). The work of E.S. on this scientific paper was supported by the Onassis Foundation — Scholarship ID: F ZP 052-1/2019-2020. The authors would like to thank Dr. Nisha Mehta (WIS) for critical reading of the manuscript.

This research was funded by the Israel Science Foundation (grant 1969/20) and by the Minerva Foundation (grant 2020/05).

The Supporting Information (in PDF format) is available free of charge at

https://doi.org/10.1021/xxxxxxx.

Appendix on the Bayesian Information Criterion; Abridged details of all 55 subsets of GMTKN55 with proper references; Optimized parameters, total WTMAD2 and its division into five major subsets for several XYG8@B3LYP variants and original XYG7@B3LYP; Total WTMAD2 and its division into five major subsets for different variants of XYG8@B*n*LYP, XYG9$_{MP3}$@BnLYP and XYG9@BnLYP; MAD, MSD and RMSD as well as breakdown of total WTMAD2 per subset for the XYG8[*f1*]@B$_{20}$LYP, XYG8[*f2*]@B$_{25}$LYP, original XYG7@B3LYP, XYG9$_{MP3}$[*f1*]@B$_{67}$LYP, XYG9$_{MP3}$[*f1*]@B$_{44}$LYP, XYG9$_{MP3}$[*f2*]@B$_{67}$LYP, XYG9[*f1*]@B$_{44}$LYP, XYG9[*f2*]@B$_{44}$LYP, XYG9[*f2g*]@B$_{44}$LYP, XYG9[*f3*]@B$_{44}$LYP, XYG9[*f5*]@B$_{44}$LYP ; and MRCC and QCHEM sample inputs for the XYG8[*f1*]@B$_{25}$LYP and XYG9[*f2*]@B$_{44}$LYP functionals.

TOC graphic (scalable):

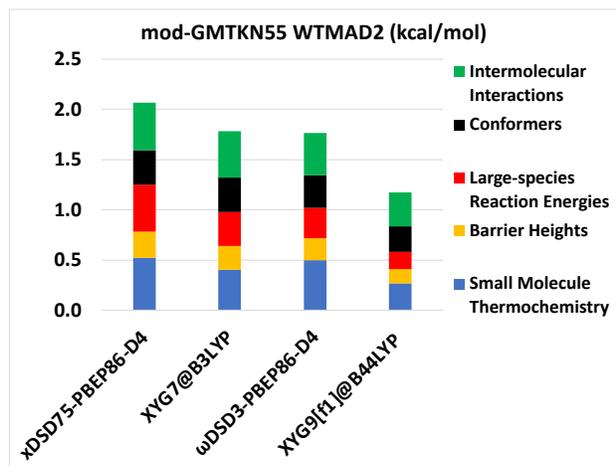